\documentclass[onecolumn,showpacs]{revtex4}
\usepackage{epsfig}
\usepackage{graphicx}%
\usepackage{amsmath}

\makeatletter
\def\btt#1{\texttt{\@backslashchar#1}}%
\DeclareRobustCommand\bblash{\btt{\@backslashchar}}%
\makeatother

\begin{document}

\title{CASIMIR PISTONS FOR MASSIVE SCALAR FIELDS}
\
\author{XIANG-HUA ZHAI}
\author{YAN-YAN ZHANG}

\author{XIN-ZHOU LI}\email{kychz@shnu.edu.cn}

\affiliation{Shanghai United Center for Astrophysics(SUCA), Shanghai
Normal University, 100 Guilin Road, Shanghai 200234,
China
}%

\date{\today}

\begin{abstract}
\noindent The Casimir force on two-dimensional pistons for massive
scalar fields with both Dirichlet and hybrid boundary conditions is
computed. The physical result is obtained by making use of
generalized $\zeta$-function regularization technique. The influence
of the mass and the position of the piston in the force is studied
graphically. The Casimir force for massive scalar field is compared
to that for massless scalar field.

\vspace{0.5cm}

\end{abstract}

\pacs{1.100}

\maketitle

About 60 years ago Casimir gave the prediction that an attractive
force should act between two plate-parallel uncharged perfectly
conducting plates in vacuum\cite{Casimir}. Especially in recent 10
years, the effect has been paid more attention because of the
development of precise measurements\cite{Decca}. At the same time,
Casimir energies and forces have been calculated theoretically in
various different configurations. Different properties of the
Casimir force (attractive or repulsive) can be obtained for
different boundary conditions and different
geometries\cite{Bordag,Gies}. For example, it has been claimed that
the Casimir energy inside rectangular cavities can be either
positive or negative depending on the ratio of the
sides\cite{Lukosz,Caruso,Li1,Li2,Edery}. But the conclusion is worth
suspecting because the calculations ignore the divergent term
associated with the boundaries and the nontrivial contribution from
the outside region of the box\cite{Barton,Graham1,Graham2}.
Recently, a modification of the rectangle$-$"Casimir piston" was
introduced to avoid the above problems\cite{Cavalcanti}. The Casimir
force on the piston is a well-defined force because the position of
the piston is independent of the divergent terms in the interval
vacuum energy and external region. Successively, the study of this
geometry attracted a lot of interests. The Casimir force on the
piston was studied for different dimensions, different fields and
different boundary
conditions\cite{Hertzberg,Barton2,Marachevsky,Edery2,Fulling,Edery3,Rodriguez}.
The results indicate that the Casimir force on the piston can be
attractive or repulsive for different cases. The repulsive Casimir
force has special importance in that it can be applied to
microelectromechamical systems (MEMS) \cite{Serry,Chan}. We
discussed Casimir pistons for a massless scalar field with hybrid
boundary conditions and obtained the repulsive Casimir force on the
piston\cite{Zhai}.

On the other hand, the Casimir effect for the massive scalar field
also studied by some authors\cite{Elizalde,Barone,Aguiar}. As is
known that the Casimir effect disappears as the mass of the field
goes to infinity since there are no more quantum fluctuations in the
limit. But the precise way the Casimir energy varies as the mass
changes is worth studying\cite{Mohideen}. In this paper,we consider
the Casimir force on the piston for the massive scalar field with
two types of boundary conditions, that is, Dirichlet and hybrid. We
obtain our physical results using $\zeta-$function regularization
technique. We discuss the influence of the mass and the ratio of the
sides graphically. The results tell us the expectable properties of
the force on the piston as is for massless scalar field and also
tell us the variation of the force as the mass changes for different
ratio of the sides.
\begin{figure}
\epsfig{file=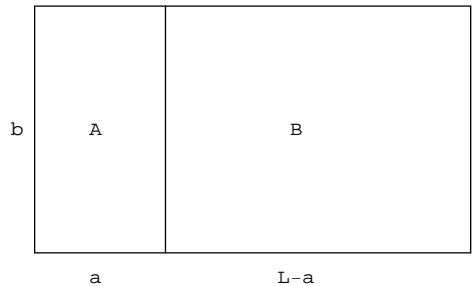,height=1.5in,width=2.5in} \caption{Casimir
piston in two dimensions.}
\end{figure}

According to \cite{Zhai}, the principal features of a 2-dimensional
piston can represent those of n-dimensional ones. Therefore, we
discuss  2-dimensional cases in detail. (see Fig.1). The piston
divides the rectangle cavity into two parts labeled $A$ and $B$ and
a quantized massive scalar field is constrained in the interval $L$.
We need to calculate the Casimir force on the piston when $L$
approaches infinity. The total energy of the vacuum for the system
can be written as the sum of three terms:
\begin{equation}
\label{eq1 }
 E=E^A(a,b)+E^B(L-a,b)+E^{out}
 \end{equation}
\noindent where $E^A(a,b)$ and $E^B(L-a,b)$ are given by the result
through cut$-$off technique, which consist of divergent terms and
finite terms, where the finite terms are the same as what are
obtained by $\zeta-$function regularization denoted $E_R^A(a,b)$ and
$E_R^B(L-a,b)$. The divergent terms and the energy from the exterior
region in the total energy are independent of the position of the
piston\cite{Cavalcanti}, so the Casimir force on the piston is as
follows:
\begin{equation}
\label{eq2} F(a)=- \frac{\partial}{\partial
a}\Big[E_R^A(a,b)+E_R^B(L-a,b)\Big]
\end{equation}

In order to calculate the Casimir energy and Casimir force, we have
to fix the boundary conditions. We consider two types of boundary
conditions. One is Dirichlet(D.B.), that is, the boundary conditions
on all surfaces are Dirichlet. The other is hybrid(H.B.), that is,
the boundary condition on the piston is Neumann, and the ones on
other surfaces are Dirichlet.

In cavity A, the vacuum energy is given by ($\hbar=c=1$ where c is
the speed of light)
\begin{widetext}
\begin{equation} \label{eq3}
 E^A(a,b)=
 \begin{cases}
\frac{1}{2}\sum_{n_1n_2=1}^\infty\Big[\Big(\frac{n_1\pi}{a}\Big)^2
+\Big(\frac{n_2\pi}{b}\Big)^2+m^2\Big]^{\frac{1}{2}},& \text{for
D.B.},\\
\frac{1}{2}\sum_{n_1n_2=1}^\infty\Big[\Big(n_1+\frac{1}{2}\Big)^2\frac{\pi^2}{a^2}
+\Big(\frac{n_2\pi}{b}\Big)^2+m^2\Big]^{\frac{1}{2}},& \text{for
H.B.}.
\end{cases}
\end{equation}
\end{widetext}
\noindent In order to use $\zeta-$function regularization, we need
to re-express Eq.(\ref{eq3}) as:
\begin{equation}
\label{eq4} E^A_R(a,b)=\begin{cases} \mathcal{E}(a,b,m), &
\text{for D.B.},\\
\mathcal{E}(2a,b,m)-\mathcal{E}(a,b,m),& \text{for H.B.}.
\end{cases}
\end{equation}
\noindent where
\begin{eqnarray}
\label{eq5}\mathcal{E}(a,b,m)
&=&\frac{1}{8}\Big\{\sum_{n_1n_2=-\infty}^\infty\Big[\Big(\frac{n_1\pi}{a}\Big)^2
+\Big(\frac{n_2\pi}{b}\Big)^2+m^2\Big]^\frac{1}{2}\nonumber\\
&-&\sum_{n_1=-\infty}^\infty\Big[\Big(\frac{n_1\pi}{a}\Big)^2+m^2\Big]^\frac{1}{2}
-\sum_{n_2=-\infty}^\infty\Big[\Big(\frac{n_2\pi}{b}\Big)^2+m^2\Big]^\frac{1}{2}-m\Big\}
\end{eqnarray}
\noindent Using Epstein $\zeta$-function\cite{Kirsten}
\begin{eqnarray}
\label{eq6} Z_p(s;a_1,\cdots,a_p;m)
&=&\sum_{n_1\cdots{n_p}=-\infty}^\infty\Big[a_1n_1^2+\cdots+a_pn_p^2+m^2\Big]^{-s}\nonumber\\
&=&\frac{\pi^{\frac{p}{2}}}{\sqrt{a_1\cdots{a_p}}}\frac{\Gamma\Big(s-\frac{p}{2}\Big)}{\Gamma\Big(s\Big)}m^{p-2s}
+\frac{\pi^s}{\sqrt{a_1\cdots{a_p}}}\frac{2}{\Gamma\Big(s\Big)}\nonumber\\
&\times&\sum_{n_1\cdots{n_p}=-\infty}^{\infty\hspace{0.3cm}\prime}m^{\frac{p}{2}-s}\Big[\frac{n_1^2}{a_1}+\cdots+
\frac{n_p^2}{a_p}\Big]^{\frac{1}{2}\Big(s-\frac{p}{2}\Big)}\nonumber\\
&\times&
K_{\frac{p}{2}-s}\bigg(2\pi{m}\Big[\frac{n_1^2}{a_1}+\cdots+\frac{n_p^2}{a_p}\Big]^\frac{1}{2}\bigg)
\end{eqnarray}
\noindent where $K_{\nu}(z)$ is the second type of modified Bessel
function and the prime on the sum denotes that the term
$n_1=n_2=\cdots=n_p$ is omitted, so we can re-express Eq.(\ref{eq5})
as
\begin{eqnarray}
\label{eq7}
 \mathcal{E}(a,b,m)
 &=&\frac{1}{8}\bigg\{\frac{m^3ab}{\pi}\frac{\Gamma(-\frac{3}{2})}{\Gamma(-\frac{1}{2})}
 +\frac{m^{\frac{3}{2}}ab}{\pi}\frac{2}{\Gamma(-\frac{1}{2})}
 \sum_{n_1n_2
 =-\infty}^{\infty\hspace{0.3cm}\prime}\frac{K_\frac{3}{2}\Big(2m\sqrt{n_1^2a^2+n_2^2b^2}\Big)}{\Big(n_1^2a^2+n_2^2b^2\Big)^\frac{3}{4}}\nonumber\\
 &-&\frac{m^2a}{\pi^\frac{1}{2}}\frac{\Gamma(-1)}{\Gamma(-\frac{1}{2})}
 -\frac{2m}{\pi^\frac{1}{2}\Gamma(-\frac{1}{2})}
 \sum_{n_1=-\infty}^{\infty\hspace{0.3cm}\prime}\frac{K_1(2man_1)}{n_1}\nonumber\\
 &-&\frac{m^2b}{\pi^\frac{1}{2}}\frac{\Gamma(-1)}{\Gamma(-\frac{1}{2})}
 -\frac{2m}
 {\pi^\frac{1}{2}\Gamma(-\frac{1}{2})}\sum_{n_2=-\infty}^{\infty\hspace{0.3cm}\prime}\frac{K_1(2mbn_2)}{n_2}-m\bigg\}
\end{eqnarray}
\noindent Instead $a$ by $L-a$, one can obtain the finite energy
$E^B_R(L-a,b)$ in cavity $B$. Substituting the expression for
$E_R^A(a,b)$ and $E_R^B(L-a,b)$ into Eq.(\ref{eq2}) and letting
$L\rightarrow \infty$, the resulting Casimir forces on the piston
$F_D(a,m)$(for D.B.) and $F_H(a,m)$ (for H.B.) are:
\begin{eqnarray}
\label{eq8} F_D(a,m)
&=&-\frac{1}{4\Gamma(-\frac{1}{2})}\bigg\{\frac{4m^\frac{3}{2}b}{\pi}\sum_{n_1n_2=1}^\infty
\frac{2K_\frac{3}{2}\Big(2m\sqrt{n_1^2a^2+n_2^2b^2}\Big)}{\Big(n_1^2a^2+n_2^2b^2\Big)^\frac{3}{4}}\nonumber\\
&-&\frac{8m^\frac{5}{2}a^2b}{\pi}\sum_{n_1n_2=1}^\infty\frac{n_1^2K_\frac{5}{2}\Big(2m\sqrt{n_1^2a^2+n_2^2b^2}\Big)}
{\Big(n_1^2a^2+n_2^2b^2\Big)^\frac{5}{4}}\nonumber\\
&+&\frac{2m^\frac{3}{2}b}{\pi{a}^\frac{3}{2}}\sum_{n_1=1}^\infty\frac{2K_\frac{3}{2}(2mn_1a)}{n_1^\frac{3}{2}}
-\frac{4m^\frac{5}{2}b}{\pi{a}^\frac{1}{2}}\sum_{n_1=1}^\infty\frac{K_\frac{5}{2}(2mn_1a)}{n_1^\frac{1}{2}}\nonumber\\
&-&\frac{2m}{\pi^\frac{1}{2}a}\sum_{n_1=1}^\infty\frac{K_1(2mn_1a)}{n_1}
+\frac{4m^2}{\pi^\frac{1}{2}}\sum_{n_1=1}^{\infty}K_2(2mn_1a)
 \bigg\}
\end{eqnarray}
 \noindent and
\begin{equation}
\label{eq9} F_H(a,m)=2F_D(2a,m)-F_D(a,m)
\end{equation}
 \noindent In order to consider the influence of the
mass, we write down the Casimir force for massless scalar field on
the piston \cite{Cavalcanti,Zhai} as follows:
\begin{equation}
\label{10}
 F_D(a,0)=-\frac{b\zeta(3)}{8\pi a^3}+\frac{\pi}{48a^2}-\frac{\zeta(3)}{16\pi
 b^2}
 +\frac{\pi b}{a^3}\sum_{n_1,n_2=1}^{\infty}n_2^2K_0\Big(2\pi n_1n_2\frac{b}{a}\Big)
 \end{equation}
\noindent and
\begin{equation}
\label{11}
 F_H(a,0)=\frac{3b\zeta(3)}{32\pi a^3}-\frac{\pi}{96a^2}-\frac{\zeta(3)}{16\pi
 b^2}+\frac{\pi b}{4a^3}\times
 \sum_{n_1,n_2=1}^{\infty}n_2^2\Big[K_0\Big(\pi
 n_1n_2\frac{b}{a}\Big)-4K_0\Big(2\pi n_1n_2\frac{b}{a}\Big)\Big]
 \end{equation}

we study the Casimir force on the piston for a massive scalar field
graphically. We can see the variation of the Casimir force depending
on the mass of the scalar field and also the influence of the ratio
of the sides $b/a$. It is worth emphasizing that compared to the
case of massless scalar field, the product $ma$ appears as a
variable. We can discuss the influence of the mass through fixing
the distance $a$.
\begin{figure}
\epsfig{file=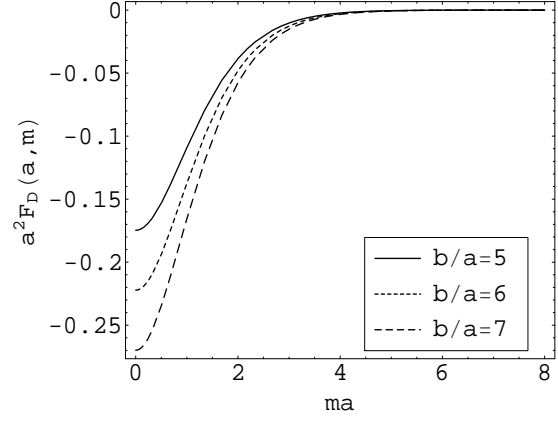,height=2.5in,width=3in} \caption{Plot of the
Casimir force on the piston (in units of $\frac{1}{a^2}$) versus
$ma$ for different ratio of $b/a$ with D.B..}
\end{figure}

\begin{figure}
\epsfig{file=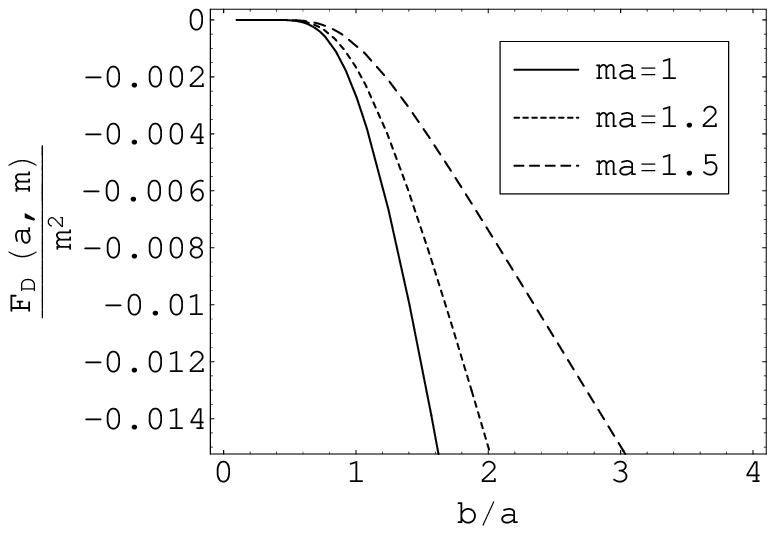,height=2.5in,width=3in} \caption{Plot of the
Casimir force on the piston (in units of $m^2$) versus $b/a$ for
different value of $ma$ for D.B..}
\end{figure}

\begin{figure}
\epsfig{file=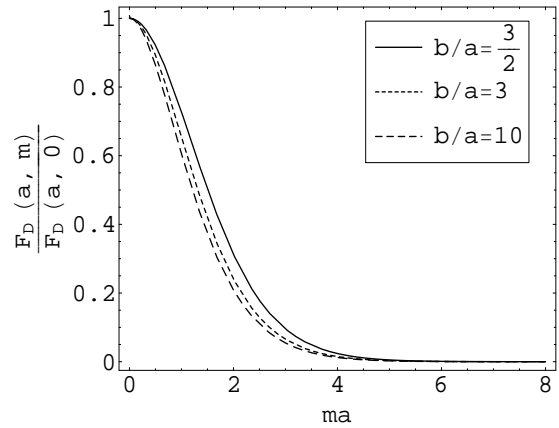,height=2.5in,width=3in} \caption{Plot of the
ratio of the Casimir force on the piston for a massive scalar field
and massless scalar field versus $ma$ for different ratio of $b/a$
for D.B..}
\end{figure}
Fig.2 to Fig. 4 are the results for D.B. and Fig.5 to Fig.7 are for
H.B.. In Fig.2 and Fig.5, choosing different ratio of the sides, we
plot the dependence of the Casimir force (in units of
$\frac{1}{a^2}$) on the mass for fixed distance. It is clear that
the force on the piston is attractive for D.B. and it is repulsive
for H.B. and it vanishes as the mass goes to infinity. In Fig.3 and
Fig.6, fixing the mass, we give the force (in units of $m^2$) with
$b/a$ increasing, where we choose different value of $ma$. The graph
tells that the force on the piston increases rapidly with the ratio
$b/a$ increasing, which is similar to the result for massless scalar
field. In order to find the influence of the mass in the Casimir
force more explicitly, we plot the ratio $F(a,m)/F(a,0)$ via the
product $ma$ in Fig.4 and Fig.7, where we plot three curves for
different ratio of $b/a$, respectively. From the graphs we can see
that the ratio approaches 1 in the limit $ma=0$ and it tends to zero
as $m$ goes to infinity. Furthermore with $b/a$ increasing the
change of the ratio with $m$ is insensitive for different $b/a$.

Our main results are summarized as follows: (i) The Casimir force on
the piston for massive scalar field is attractive for D.B. and it is
repulsive for H.B., which is the same result as that for massless
scalar field. (ii) The Casimir force decreases with the mass
increasing. The force vanishes as the mass goes to infinity, while
in the limit $m=0$ the force recovers the result for massless scalar
field . (iii) For fixed mass the force on the piston increases as
the piston moves close to the nearest end, which is also similar to
the result for massless scalar field.
\begin{figure}
\epsfig{file=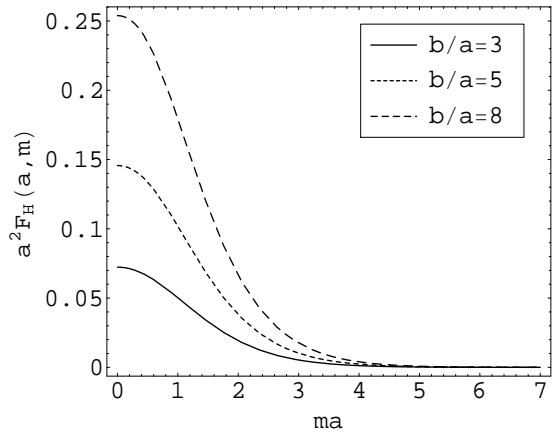,height=2.5in,width=3in} \caption{Plot of the
Casimir force on the piston (in units of $\frac{1}{a^2}$) versus
$ma$ for different ratio of $b/a$ with H.B..}
\end{figure}

\begin{figure}
\epsfig{file=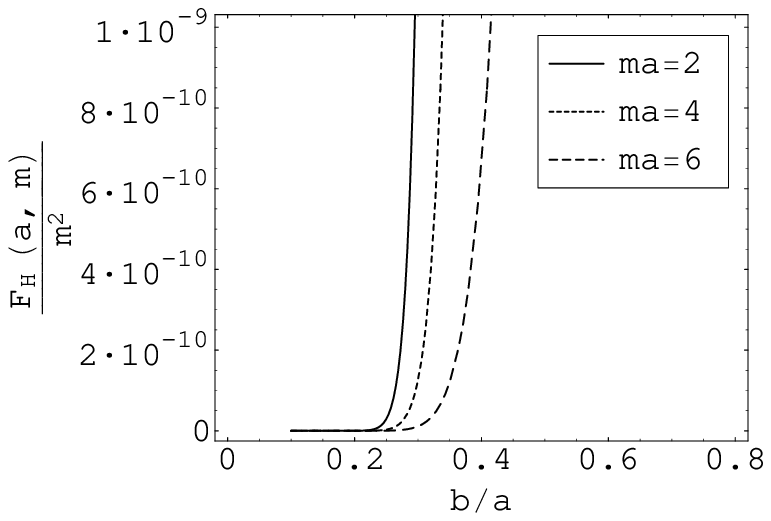,height=2.5in,width=3in} \caption{Plot of the
Casimir force on the piston (in units of $m^2$) versus $b/a$ for
different value of $ma$ for H.B..}
\end{figure}

\begin{figure}
\epsfig{file=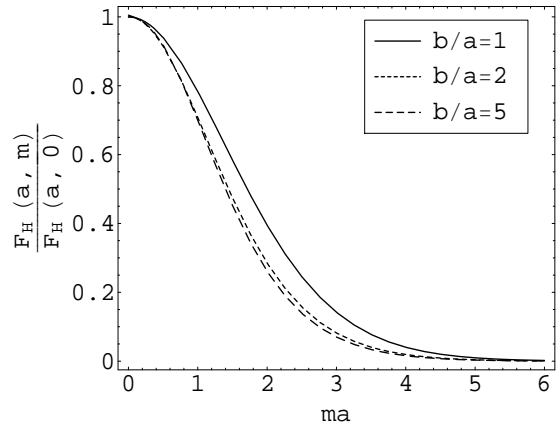,height=2.5in,width=3in} \caption{Plot of the
ratio of the Casimir force on the piston for a massive scalar field
and massless scalar field versus $ma$for different ratio of $b/a$
for H.B..}
\end{figure}
For three-dimensional pistons, one can discuss similarly and can get
the same result but with more complicated calculations.

\vspace{0.8cm} \noindent ACKNOWLEDGEMENT: This work is supported by
National Nature Science Foundation of China under Grant No. 10671128
and Shanghai Municipal Education Commission(No 06DZ005).

\end{document}